\documentclass[preprint2]{aastex}

\shorttitle{HBC 425}
\shortauthors{Dahm and Lyke}

\begin{document}

\title{The Low-Mass Companion to the Lithium-Depleted, Spectroscopic Binary HBC 425 (St 34)}

\author{S. E. Dahm\altaffilmark{1} and J. E. Lyke\altaffilmark{1}}

\altaffiltext{1}{W. M. Keck Observatory, Kamuela, HI 96743}

\begin{abstract}
We present high angular resolution, near-infrared imaging and spectroscopy of a low-mass
companion to the lithium-depleted, double-line spectroscopic binary HBC 425 (St 34) obtained
using the Near Infrared Spectrograph (NIRSPEC) and the Keck II adaptive optics system. 
Positioned $\sim1\farcs23$ southeast of the primary pair, the companion, HBC 425C, is 
$\sim$2.4 magnitudes fainter at 2.2 $\mu$m. Moderate-resolution ($R\sim2500$) $J-$ and 
$K-$band spectroscopy reveal HBC 425C to have an M5.5 ($\pm$0.5) spectral type. Comparisons 
with pre-main sequence evolutionary models imply a mass of $\sim$0.09 M$_{\odot}$ and
ages of 8--10 Myr, assuming the nominal distance of Taurus-Auriga ($\sim$140 pc), or
$\sim$25 Myr if placed at $\sim$90 pc. We also present high dispersion, optical spectra of 
HBC 425 and HBC 425C obtained using the High Resolution Echelle Spectrometer (HIRES) on 
Keck I. We detect strong \ion{Li}{1} $\lambda$6708 absorption in the spectrum of HBC 425C.
Using curves of growth for the \ion{Li}{1} $\lambda$6708 doublet, we estimate its abundance
level to lie between log $N$(Li)=1.9 and 3.1 dex. The spectrum of
HBC 425 exhibits \ion{Ca}{2} H \& K, \ion{He}{1} $\lambda\lambda$5876, 6678, and strong
Balmer line emission, consistent with accretion. We place more restrictive upper limits on the surface
abundance of lithium and find that HBC 425 retains less than $\sim$0.1\% of its primordial
abundance. The presence of lithium in the photosphere of HBC 425C does not resolve the 
discrepancy between isochronal and lithium depletion ages for the primary pair. However,
if lithium were depleted relative to interstellar abundance levels, even minimally, 
considerable support would be gained for the more advanced age of this hierarchical triple system.
\end{abstract}

\keywords{binaries: spectroscopic --- stars: abundances --- stars: pre-main sequence}

\section{Introduction}
Low-mass ($M\le$1 M$_{\odot}$), pre-main sequence stars are fully convective as they
gravitationally contract and descend their vertical tracks in the color-magnitude diagram 
(Hayashi 1961). When core temperatures reach 2.5 to 3.0 MK, lithium depletion begins as 
its dominant isotope ($^{7}$Li) fuses with hydrogen nuclei to form $^{4}$He (Bodenheimer 1965). 
Convective processes circulate material through the stellar core, depleting all available 
lithium over timescales of tens of Myr (Baraffe et al. 1998; Song et al. 2002, and 
references therein). In spite of its low abundance, $^{7}$Li is readily observed in 
stellar atmospheres in an electronic transition that is a strong doublet at 
$\lambda\lambda$6707.78, 6707.93. Given the short timescale of lithium depletion, 
the presence of \ion{Li}{1} $\lambda$6708 absorption in the spectra of low-mass 
stars is recognized as an indicator of youth (e.g. Bonsack \& Greenstein 1960; Herbig 1965). 
Stars presumably begin their lives with an amount of lithium consistent with 
interstellar abundance levels. Anders \& Grevesse (1989) derive an abundance of lithium 
extracted from carbonaceous chondrites of $N$(Li)=3.31$\pm$0.04 dex, which is adopted here as 
the interstellar $^{7}$Li abundance level.

HBC 425 (St 34) was first identified as an H$\alpha$ emission star in the objective
prism survey of Stephenson (1986). The star was classified as an early to mid M-type 
star exhibiting Balmer line and \ion{He}{1} $\lambda$5876 emission by Downes \& Keyes (1988).
The spectral type was further refined by White \& Hillenbrand (2005) to M3, who
determined that HBC 425 is a double-line spectroscopic binary with components
of identical effective temperature and luminosity. HBC 425 lies $\sim$1.5$^{\circ}$ 
from the molecular cloud core L1558 near the southern extremity of the Taurus-Auriga 
molecular cloud complex and was subsequently assumed to be a member of the star forming 
region given its strong T Tauri-like attributes.

Surprisingly, White \& Hillenbrand (2005) discovered that neither component of the double-line
spectroscopic binary exhibits \ion{Li}{1} $\lambda$6708 absorption. The isochronal 
age of HBC 425 was determined to be 8$\pm$3 Myr, substantially younger than the predicted 
lithium depletion timescale of $\sim$25 Myr using the pre-main sequence models of 
Baraffe et al. (1998). White \& Hillenbrand (2005) suggested that a distance of 
$\sim$90 pc would reconcile the discrepancy between isochronal and lithium depletion 
ages at $\sim$21-25 Myr. Hartmann et al. (2005) favor this interpretation making
HBC 425 the oldest known accreting pre-main sequence star. Given such strong 
evidence for membership in Taurus-Auriga, however, White \& Hillenbrand (2005) conclude
that HBC 425 was likely much younger than implied by lithium depletion.

Hartmann et al. (2005) obtained a mid-infrared spectrum of HBC 425 using the Infrared 
Spectrograph (IRS) on {\it Spitzer Space Telescope} (Werner et al. 2004). The spectrum 
revealed substantial excess emission at wavelengths $>$5.4 $\mu$m, but the lack of excess 
emission at shorter wavelengths (i.e. 2.2 $\mu$m) implied the presence of an evacuated 
inner disk region. The depleted inner disk would likely result from dynamical clearing 
by the two components of the spectroscopic binary, similar to the CoKu Tau 4 system that 
was previously recognized as a transition disk source (Ireland \& Kraus 2008). 
Andrews \& Williams (2005) place an upper limit on the mass of the circumbinary disk of 
$<$5$\times$10$^{-4}$ M$_{\odot}$.

The kinematic survey of Taurus-Auriga by Bertout \& Genova (2006) found insignificant 
proper motions for HBC 425 as well as for 10 other assumed Taurus-Auriga members including
CoKu Tau 4, DQ Tau, and V836 Tau. While purely circumstantial in nature, the lack of 
significant proper motion may argue in favor of a larger distance estimate for HBC 425, 
i.e. more consistent with the distance of Taurus-Auriga.

In this work we present near-infrared, adaptive optics imaging and spectroscopy of
HBC 425C, an apparent low-mass companion positioned $\sim1\farcs23$ southeast of the 
spectroscopic binary. This component was first noted by White \& Hillenbrand (2005)
who estimated it to be $\sim$2.5 magnitudes fainter than the primary in $K-$band. 
We use the pre-main sequence models of Baraffe et al. (1998) to re-evaluate the 
mass and age of the HBC 425 system. We also present high dispersion optical spectra
of HBC 425 and HBC 425C and use these spectra to examine the kinematics of the system
and to place limits on the abundance of lithium present in their photospheres.

\section{Observations}

\subsection{Adaptive Optics Near Infrared Imaging and Spectroscopy}
The Near Infrared Spectrograph (NIRSPEC; McLean et al. 1998) installed behind the adaptive 
optics system of Keck II (NIRSPAO) was used on 2010 December 12 to obtain moderate-dispersion 
($R\sim2500$) $J$ and $K-$ band spectra of HBC 425 and HBC 425C. The spectrograph was configured
with the 3-pixel slit and the NIRSPEC-3 ($J-$band) and NIRSPEC-7 ($K-$band) filters. 
Using the rotator of the adaptive optics system, the slit was aligned with the axis
of the visual binary (position angle $\sim$137$^{\circ}$.9). The stars were nodded along 
the slit and observed in 3 positions to ensure proper sky subtraction and reasonable
signal to noise levels in the extracted spectra. Integration times were 180 and 120 s 
in the NIRSPEC-3 and NIRSPEC-7 filters, respectively. Flat lamp and argon arcline spectra 
were obtained for each instrument configuration to remove pixel to pixel variations and for 
wavelength calibration. A nearby, bright A0 V star (HD 35036) was observed immediately after 
HBC 425 in each filter for telluric correction. Basic image reduction, spectral extraction,  
and wavelength calibration were achieved using REDSPEC, an analysis package written by L. Prato, 
S. Kim, and I. McLean that is publically available through the NIRSPEC webpage.

The slit viewing camera (SCAM) was used to image HBC 425 and HBC 425C in the NIRSPEC-3, 
5, and 7 filters ($J,H,K-$band, respectively). Behind the adaptive optics system, the 
platescale of SCAM is 0.0168 arcsec per pixel, yielding a nominal field of view of 
4$\farcs$3 square. Integration times per coadd were 1 s or less to maintain linearity,
requiring the use of correlated double sampling (CDS) readout mode. Detector read noise 
in CDS mode is $\sim$10 e$^{-1}$. The field was dithered with HBC 425 and HBC 425C off the 
slit in several positions to allow for sky and dark current subtraction. Aperture 
photometry was performed on HBC 425 and HBC 425C using the {\it phot} task in the 
DAOPHOT package of IRAF on sky-subtracted images. An aperture radius of 10 pixels 
(0$\farcs$168) was adopted for the analysis. Photometric uncertainties are represented by 
the standard deviation of measurements on several sky-subtracted frames in each filter. 
Absolute photometric calibration was achieved by using the 2MASS $J$, $H$, and $K_{S}-$band 
magnitudes for HBC 425. Photometry for these sources is presented in Table 1.

\subsection{High Dispersion Optical Spectroscopy}
The High Resolution Echelle Spectrometer (HIRES) is a grating cross-dispersed 
spectrograph permanently mounted on the Nasmyth platform of Keck I (Vogt et al. 1994). 
High dispersion spectra were obtained using HIRES with the red cross-disperser and 
collimator in beam on the nights of 2010 December 02, 2011 October 12, and 2011 October 19 (UT),
abbreviated henceforth by J2010, J2011A, and J2011B, respectively. All nights were photometric 
with seeing conditions of $\sim$0$\farcs$85 for J2010 and J2011B and $\sim$0$\farcs$60 for 
J2011A. The B2 decker (0$\farcs$574$\times$7$\farcs$0) was used during J2010 to 
provide a spectral resolution of $\sim$66,000 ($\sim$4.5 km s$^{-1}$). Near complete 
wavelength coverage from $\sim$3650--7900\AA\ was achieved, a region that includes 
many gravity and temperature sensitive photospheric features as well as permitted 
and forbidden transitions generally associated with accretion or chromospheric activity: 
\ion{Ca}{2} H \& K, H$\gamma$, H$\beta$, \ion{He}{1} $\lambda$5876, [O I] $\lambda$6300, 
and H$\alpha$. For the J2011 observations, the cross-disperser angle was altered to 
provide expanded wavelength coverage in the red, from $\sim$4300--8680\AA\ . The C1 
and C5 deckers (0$\farcs$861$\times$7$\farcs$0 and 1$\farcs$148$\times$7$\farcs$0) 
were used to provide spectral resolutions of $\sim$50,000 (6.0 km s$^{-1}$) and 
37,500 (8.0 km s$^{-1}$), respectively. The excellent seeing conditions on 2011 October 12
allowed the placement of the slit directly on HBC 425C, perpendicular to the axis 
joining HBC425C to HBC 425. The one hour integration yielded a signal to noise level
of $\sim$20 near $\lambda$6700.

The 3-chip mosaic of MIT-LL CCDs with 15 $\mu$m pixels was used in low gain mode 
resulting in readout noise levels of $\sim$2.8, 3.1, and 3.1 e$^{-1}$, for the red, 
green, and blue detectors, respectively. Internal quartz lamps were used for flat 
fielding and ThAr lamp spectra were obtained for wavelength calibration. Several
radial and rotational velocity standard stars having M0--M3.5 spectral types from
Nidever et al. (2002) and Browning et al. (2010) were observed during the nights.
The HIRES observations were reduced using the Mauna Kea Echelle Extraction, {\it makee}, 
reduction script written by Tom Barlow, which is publically available through links on 
the HIRES webpage.

\section{HBC 425: The Spectroscopic Binary}

The heliocentric radial velocities of the individual components of HBC 425 were determined
by cross-correlation analysis using the M0.5 type radial velocity standard HD 28343
as a template. To distinguish primary from secondary, we assume that the depths of absorption 
lines are greater for the former than for the latter, implying a slight difference in
spectral type. The J2010 HIRES spectrum of HBC 425 was obtained when the velocities
of the two components were nearly reversed from those reported by White \& Hillenbrand (2005),
$-$10.67 and +46.32 km s$^{-1}$ for the primary and secondary, respectively. The J2011A
HIRES spectrum reveals single component absorption line profiles, implying equal velocities
of $\sim$+16.8 km s$^{-1}$. The J2011B radial velocities were measured to be +41.42 and 
$-$5.53 km s$^{-1}$ for the primary and secondary, respectively.

Lacking sufficient observations for a complete orbit determination, we use the method of 
Wilson (1941) to derive the mass ratio ($q$) and systemic velocity ($\gamma$) of the system 
defined by:

\begin{equation}
q = \frac{v_{2}-v_{1}}{u_{1}-u_{2}}
\end{equation}

\noindent and 

\begin{equation}
\gamma = \frac{u_{1}v_{2}-u_{2}v_{1}}{(v_{2}-v_{1})-(u_{2}-u_{1})}
\end{equation}

\noindent where $u_{1}$ and $u_{2}$ are the J2010 and J2011B velocities of the primary and $v_{1}$ 
and $v_{2}$ are those of the secondary. We find $q=1.00$ and $\gamma$=17.89 km s$^{-1}$.
This systemic velocity is consistent 
with that derived by White \& Hillenbrand (2005) and with the radial velocities of known 
Taurus-Auriga members.

Adopting the masses for the primary and secondary derived by White \& Hillenbrand (2005),
0.37 M$_{\odot}$, and assuming circular orbits with an inclination angle of 90$^{\circ}$,
an upper limit for the semi-major axis of the system is $\sim$0.2 AU. An upper limit for 
the period of the spectroscopic binary follows, $\sim$0.15 years or $\sim$54 days. It is
likely, however, that these observations and that of White \& Hillenbrand (2005) were not made
at maximum velocity separation and that the system is not observed edge-on. We provide a
summary of observed properties of the spectroscopic binary in Table 2 and inferred or
derived properties in Table 3.

The HIRES spectra of HBC 425 reveal strong, broad H$\alpha$ emission with W(H$\alpha$)=$-$27.4 \AA\
(J2010), $-$40.0 \AA\ (J2011A), and $-$36.8 \AA\ (J2011B). Taken together with the equivalent widths reported by 
White \& Hillenbrand (2005), W(H$\alpha$)=$-$51.6 \AA\, and Downes \& Keyes (1988), $-$78 \AA\ , 
these measurements imply significant variability. The J2010 and J2011 H$\alpha$ emission profiles are double-peaked 
and similar in appearance with velocity widths of $\sim$366 km s$^{-1}$ (J2010), 
462 km s$^{-1}$ (J2011A), and 468 km s$^{-1}$ (J2011B), indicative of accretion using the H$\alpha$ 10\% width of peak 
emission criterion of White \& Basri (2003). Higher order Balmer transitions (H$\beta$, 
H$\gamma$, H$\delta$, H$\epsilon$) are also double-peaked, but with smaller velocity separations. 
Broad, blue-shifted wings are evident in these emission lines suggestive of accretion driven 
winds. [O I] $\lambda$6300 is also found in emission with radial velocities of +10.6 km s$^{-1}$
(J2010), +18.1 km s$^{-1}$ (J2011A), and +20.2 km s$^{-1}$ (J2011B). The uncertainties for these 
velocities are $\pm$2 km s$^{-1}$. Such forbidden emission may arise from a circumbinary disk wind
given its near systemic velocity (e.g. Hartigan et al. 1995).
There is no evidence of [S II] $\lambda\lambda$6717, 6730 emission in the HIRES spectra.

The \ion{Ca}{2} H \& K lines (J2010) are double-peaked, but while strongly in emission the features
are narrow and lack the broadened wings evident in the Balmer lines. These \ion{Ca}{2} emission
peaks are consistent with the velocities of the primary and secondary, implying that they arise
from chromospheric activity associated with each component. The \ion{Na}{1} D lines (J2010) exhibit
central absorption flanked by emission peaks that also correlate with the velocities of the
primary and secondary. The strength of these chromospheric emission lines (i.e. \ion{Ca}{2},
\ion{Na}{1}) is greater in the secondary than in the primary. The \ion{Ca}{2} near-infrared
triplet ($\lambda$$\lambda$8498, 8542, 8662) was included in the J2011 spectra. These features
exhibit emission reversal that nearly fills the wings of the underlying absorption profiles.
The radial velocities of the double emission lines present in the \ion{Ca}{2} near-infrared 
triplet in the J2011B spectrum again correlate with the velocities of the primary and secondary.
Shown in Figure 1 are sections of the J2010, J2011A, and J2011B HIRES spectra including \ion{Ca}{2} K, 
\ion{Ca}{2} $\lambda$8542, \ion{Ca}{1} $\lambda$6102.72 (exemplifying the double-lined features 
observed in the J2010 and J2011B spectra), and H$\alpha$. 

Lithium is clearly depleted in both components of the spectroscopic binary. Shown in Figure 2 is 
an expanded region of one order of the J2010 spectrum centered near \ion{Li}{1} $\lambda$6708. 
The spectrum of the M3-type main sequence star Gl 806 is shown for comparison. Although weak 
absorption features appear at the wavelengths expected for the primary and secondary, the 
strengths of these features are consistent with the level of variation observed in the pseudo-continuum 
of the star. The measured equivalent widths, $\sim$30 m\AA\, serve as upper limits that are a 
factor of two lower than that determined by White \& Hillenbrand (2005). Using the curves 
of growth for the \ion{Li}{1} $\lambda$6708 transition derived by Pavlenko \& Magazzu (1996) 
and adopting an effective temperature of 3500 K with log $g$=4.5, we determine an upper limit 
for the surface \ion{Li}{1} abundance in HBC 425 to be between log $N$(Li)=$-$0.5 and $-$1.0 dex. 
From this we infer that lithium is depleted in the photospheres of HBC 425A+B by at least 
$\sim$3 orders of magnitude relative to interstellar abundance levels.

\section{HBC 425C: The Late-Type, Wide Binary Companion}

HIRES guide camera images (Figure 3a) revealed the presence of a tertiary component of HBC 425.
White \& Hillenbrand (2005) noted this companion and estimated it to be $\sim$2.5 magnitudes 
fainter in $K-$band than the primary. The companion is separated by $\sim$1$\farcs23$, corresponding to a 
projected separation of $\sim$172 AU assuming the traditionally accepted distance of Taurus-Auriga,
($\sim$140$\pm$10 pc: Kenyon et al. 1994). To estimate the probability of a chance alignment, we follow the example
of Metchev \& Hillenbrand (2009) and sum the number of sources within a 5\arcmin\ radius of 
HBC 425 in the 2MASS point source catalog. Multiplying this value by the ratio of solid angles,
(1$\farcs$23)$^{2}$ / (5\arcmin$)^{2}$, we derive a probability for a purely geometric
alignment of $\sim$0.2\%. The radial velocity of HBC 425C is also measured to be +15.9 km s$^{-1}$,
consistent with the systemic velocity of HBC 425.

Shown in Figure 3b is a background-subtracted NIRSPAO $H-$band image of HBC 425
and HBC 425C obtained using SCAM. The sources are clearly separated allowing the slit to be 
aligned along the axis of the binary without concern for overlapping point spread functions.
Shown in Figures 4 and 5 are the extracted $J-$ and $K-$band spectra respectively for HBC 425 
and HBC 425C with critical atomic and molecular features identified. Superposed and plotted with 
additive constants are spectra for standard stars (M3--M8 types) obtained from the IRTF Spectral 
library (Rayner et al. 2009). Weak Pa$\beta$ and Br$\gamma$ emission are observed in the spectra
of HBC 425, consistent with accretion activity. The strength of atomic lines and molecular features 
in the spectra of HBC 425 support the M3 classification of White \& Hillenbrand (2005). 

The $J-$ and $K-$band spectra of HBC 425C and its near-infrared colors are consistent with a 
M5.5$\pm$0.5 spectral type. The lack of strong \ion{Ca}{1} absorption near 2.26 $\mu$m 
is suggestive of being later than M5. Although the $J-$band spectrum is affected by the lower 
throughput of the N3 filter at shorter wavelengths, the \ion{K}{1} doublet near 1.17 $\mu$m is 
clearly stronger in HBC 425C than in HBC 425 and is suggestive of $\ge$M5 spectral type.
There is no evidence for Pa$\beta$ or Br$\gamma$ emission 
in the spectrum of HBC 425C, implying that this source is either not accreting or accreting at 
levels below the detection threshold for these diagnostics. The near-infrared colors of HBC 425C 
are also consistent with purely photospheric emission. It is evident from weak $J-H$ excesses 
that both HBC 425 and HBC 425C suffer minimal foreground extinction.

The HIRES spectrum of HBC 425C reveals strong \ion{Li}{1} $\lambda$6708 absorption.
Shown in Figure 2 are the spectra of HBC 425C
and UX Tau C, an M5-type Taurus-Auriga member, centered near $\lambda$6708. The measured 
equivalent width of \ion{Li}{1} $\lambda$6708, W(Li), for HBC 425C is $\sim$0.32$\pm$0.05 \AA\ , 
substantially lower than that of UX Tau C, W(Li)=0.6 \AA\ . This is in part due to scattered 
light from HBC 425 that artificially elevates the observed continuum level. To estimate the
level of scattered light contamination, we use the slit guide camera images of the HBC 425 system
taken through the RG610 photometric filter. Using the nearby field star J04542362$+$1709434 
to construct a model point spread function (psf), we remove the psf of HBC 425 from the image
and measure the flux in a circular aperture equivalent in size to the width of the C5 decker
centered on HBC 425C. Performing aperture photometry for HBC 425C in the orginal image, we 
find that $\sim$20--40\% of the incident light likely originates from HBC 425. Removing this scattered
light contribution from the continuum near $\lambda$6700 in the HIRES spectrum of HBC 425C,
we measure W(Li)=0.47$^{+0.08}_{-0.06}$ \AA\ . Using the LTE curves of growth for the
\ion{Li}{1} $\lambda$6708 transition from Zapatero Osorio et al. (2002) and assuming an effective 
temperature consistent with HBC 425C (2900--3100 K), we find that the measured equivalent
width, W(Li)$\sim$0.32 \AA\ , implies a lithium abundance of log $N$(Li)$\sim$1.6 dex, suggestive 
of substantial depletion relative to intersteller levels. Correcting W(Li) for the scattered light contribution
from HBC 425, we estimate the lithium abundance level to be between log $N$(Li)$\sim$1.9 and 3.1 dex, 
where the latter is assumed by  Zapatero Osorio et al. (2002) to be representative of the
cosmic or primordial abundance level. While it is possible that the abundance of lithium in HBC 425C 
is depleted relative to interstellar levels, we cannot conclusively demonstrate this with the
present seeing limited observations.

\section{Discussion and Conclusions}

Binaries play a crucial role in assessing the accuracy of pre-main sequence evolutionary models 
and isochrones (e.g. Hillenbrand \& White 2004, Mathieu et al. 2007). Using the effective temperature 
scale of Luhman et al. (2003), extinction corrected $J-$band magnitudes, and the main sequence bolometric corrections of Kenyon \& Hartmann (1995), 
we place HBC 425A+B and HBC 425C on the Hertzsprung-Rusell (HR) diagram shown in Figure 6a). 
Superimposed are the 1, 3, 10, 20 Myr, and 1 Gyr isochrones, and the 0.35, 0.1, and 0.08 M$_{\odot}$ 
evolutionary tracks of Baraffe et al. (1998), placed at the nominal distance of Taurus-Auriga 
$\sim$140 pc. The solar metallicity Baraffe et al. (1998) models incorporate non-grey atmospheres 
and assume a general mixing length parameter given by $\alpha$=1. Each component of HBC 425 falls near 
the 0.35 M$_{\odot}$ evolutionary track at an age of $\sim$10 Myr. HBC 425C lies near the 
0.09 M$_{\odot}$ evolutionary track with a coeval age of $\sim$8--10 Myr. 

Also depicted in Figure 6a) are the theoretical lithium depletion curves of Chabrier \& Baraffe (1997).
The onset of depletion is represented by the enhanced dashed curve and the cross-hatched 
region represents significant depletion of $\sim$3 orders of magnitude or greater from
initial abundance levels. The models predict that HBC 425 should have experienced some 
degree of lithium depletion at its assumed age of $\sim$10 Myr while HBC 425C should still
retain its primordial lithium abundance. As originally noted by White \& Hillenbrand (2005), if a 
member of Taurus-Auriga, the level of lithium depletion observed in HBC 425 is clearly 
inconsistent with the models of Chabrier \& Baraffe (1997). White \& Hillenbrand (2005) 
summarize 3 possible explanations: 1) the effective temperature of HBC 425 is $\sim$340 K 
hotter than anticipated for its M3 spectral type; 2) HBC 425 lies in the foreground of 
Taurus-Auriga by some $\sim$50 pc and is consequently much older than previously assumed; 
or 3) a problem exists in models of lithium depletion for low-mass, pre-main sequence stars. 
We can only address the second of these hypotheses with the observations presented here.

Shown in Figure 6b) is the HR diagram of the HBC 425 system placed at a distance of 90 pc.
The ages of HBC 425 and HBC 425C are coeval near $\sim$25 Myr. From their placement relative
to the region of significant lithium depletion, both components of the spectroscopic binary 
should have experienced near complete lithium destruction, as observed. The substantially 
lower mass HBC 425C, however, lies on or near the boundary representing the onset of lithium 
depletion. If shown to be lithium depleted relative to interstellar abundance levels, even 
minimally, this would greatly strengthen the argument for the more advanced age ($\sim$25 Myr)
and the more proximal distance ($\sim$90 pc) to the system.

The presence of mid-infrared excess indicative of a primordial disk around HBC 425 
(Hartmann et al. 2005) as well as strong H$\alpha$ emission consistent with accretion 
(White \& Hillenbrand 2005 and this investigation) complicates this issue considerably.
The timescale of disk dissipation within the terrestrial region has been reasonably 
well-established by ground-based (e.g. Haisch et al. 2001; Mamajek et al. 2004) and 
{\it Spitzer Space Telescope} (e.g. Uchida et al. 2004; Silverstone et al. 2006) 
observations to be $\le$10 Myr. The existence of an accretion disk having an age 
$\sim$2.5 times this limit, while certainly not beyond reason, is exceptional nonetheless.
The question of lithium depletion in HBC 425 remains unresolved. It is, however, not 
without precedent: e.g. HIP 112312A in the $\sim$12 Myr old $\beta$ Pictoris moving group 
(Song et al. 2002). While the lithium depletion age of HIP 112312A is $\sim$35 Myr,
its isochronal age is $\sim$6 Myr. High precision proper motions of HBC 425 that could 
unambiguously determine its membership status in Taurus-Auriga are needed.
Clearly a dedicated effort must be made to determine the distance to this enigmatic system 
and thereby resolve the lithium depletion problem. 

\acknowledgments
The Digitized Sky Surveys, which were produced at the Space Telescope Science Institute under 
U.S. Government grant NAG W-2166, were used as were the the SIMBAD database operated at CDS,
Strasbourg, France, and the Two Micron All Sky Survey (2MASS), a joint project of the University 
of Massachusetts and the Infrared Processing and Analysis Center (IPAC)/California Institute of 
Technology, funded by NASA and the National Science Foundation. SED is grateful to Russel White,
Lynne Hillenbrand, and G. H. Herbig for insightful discussions regarding the nature of HBC 425
and to an anonymous referee whose comments and suggestions greatly improved this manuscript.
SED is also grateful to the Director of W. M. Keck Observatory for the use of Director's time in
carrying out this program and to Heather Hershley and Terry Stickel for their dedicated efforts
in supporting these observations.
{\it Facilities:} \facility{W. M. Keck Observatory}.

\clearpage

%% the file, uncomment the \plotone, \plottwo, and
%% If you need a layout that cannot be achieved with \plotone or
%% \plottwo, you can invoke the graphicx package directly with the
%% \includegraphics command or use \plotfiddle. For more information,
%% please see the tutorial on "Using Electronic Art with AASTeX" in the
%% documentation section at the AASTeX Web site,
%% http://www.journals.uchicago.edu/AAS/AASTeX.
%%
%% The examples below also include sample markup for submission of
%% supplemental electronic materials. As always, be sure to check
%% the instructions to authors for the journal you are submitting to
%% for specific submissions guidelines as they vary from

%% This example uses \plotone to include an EPS file scaled to
%% 80% of its natural size with \epsscale. Its caption
%% has been written to indicate that additional figure parts will be
%% available in the electronic journal.

\begin{figure}
\epsscale{1.50}
\plotone{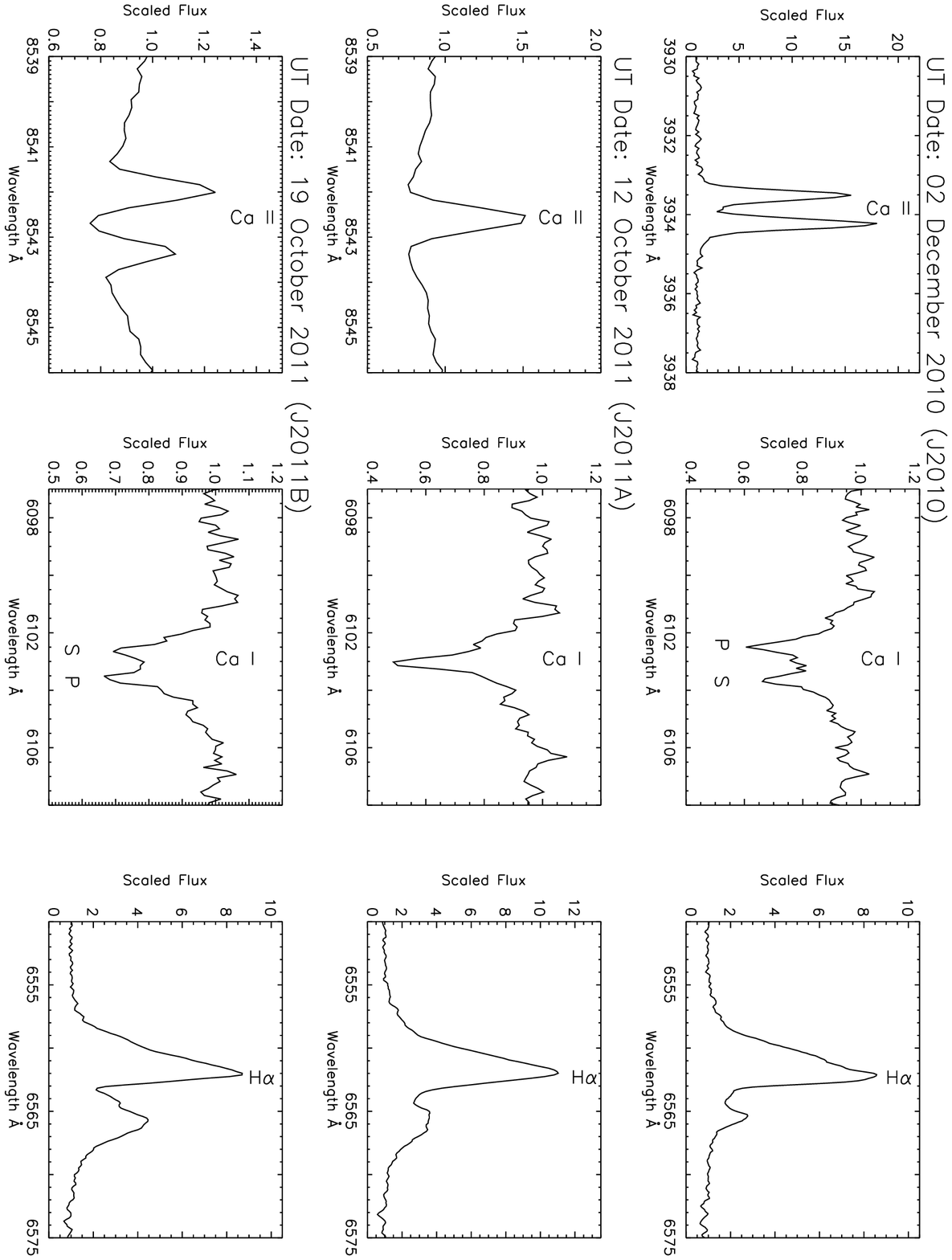}
\caption{Sections of HIRES spectra of HBC 425 obtained on (UT) 02 December 2010 (upper panels),
12 October 2011 (middle panels), and 19 October 2011 (lower panels) including: Ca II K 
exhibiting two emission peaks corresponding to the radial velocities of the primary and secondary,
Ca II $\lambda$8542, the photospheric Ca I absorption feature near $\lambda$6102 demonstrating 
the separation of the doubled lines produced by the primary and secondary, labeled $P$ and $S$,
respectively, and the strong H$\alpha$ emission profiles that are consistent with accretion.
\label{fig1}}
\end{figure}

\clearpage

\begin{figure}
\epsscale{1.50}
\plotone{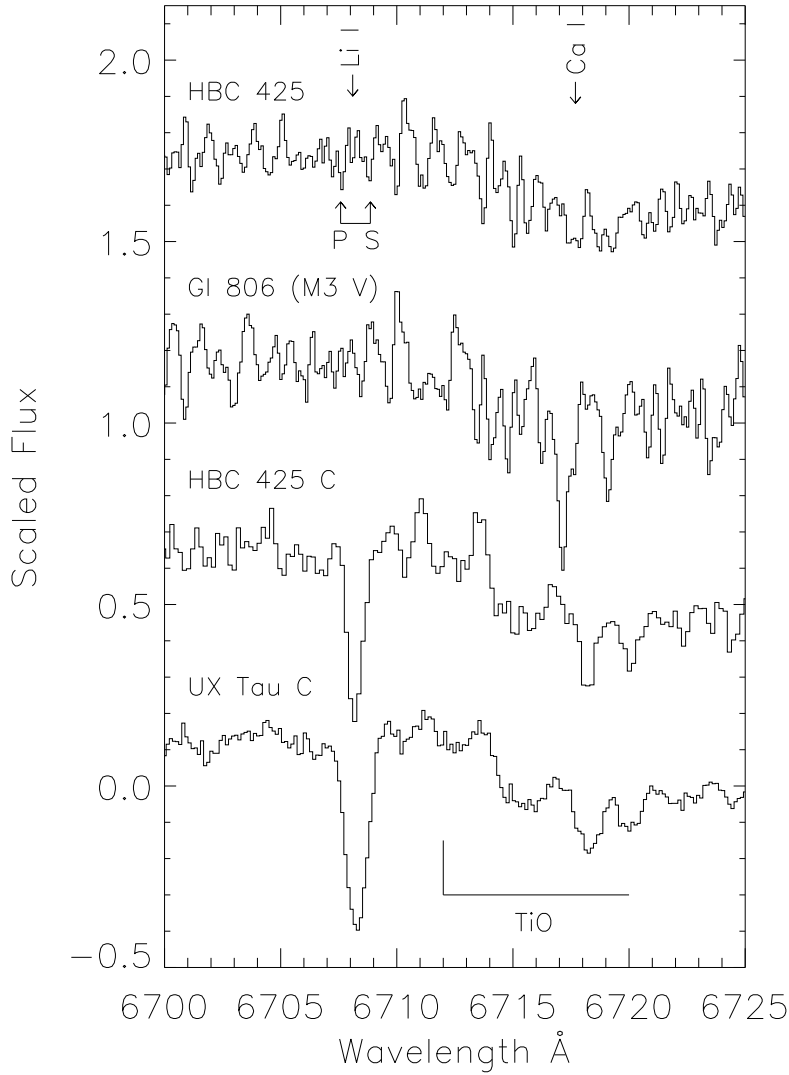}
\caption{HIRES spectra centered near  Li I $\lambda$6708 for HBC 425, the M3-type dwarf Gl 806, 
HBC 425C, and UX Tau C an M5-type member of Taurus-Auriga. The signal-to-noise level of the 
HBC 425 and HBC 425C spectra near $\lambda$6700 are $\sim$60 and $\sim$20, respectively. 
The expected positions of Li I $\lambda$6708 for the primary and secondary of HBC 425 are 
indicated by the $P$ and $S$. The features present at these positions have equivalent widths 
that are consistent with those found in the pseudo-continuum. Strong Li I $\lambda$6708 
absorption is detected in HBC 425C, W(Li)=0.32$\pm$0.05 \AA\ . Although substantially weaker than 
equivalent widths measured for UX Tau C or other late-type, Taurus-Auriga members, scattered light 
from HBC 425 contaminates the continuum. \label{fig2}}
\end{figure}

\clearpage

\begin{figure}
\epsscale{1.50}
\plottwo{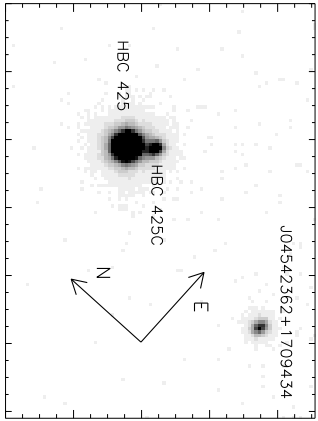}{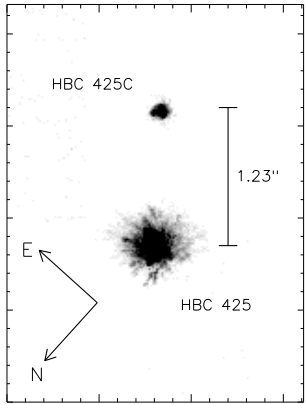}
\caption{HIRES guide camera image of HBC 425 revealing the tertiary member, HBC 425C (top). 
NIRSPAO $H-$band image of HBC 425 and HBC 425C obtained with SCAM. HBC 425 and HBC 425C 
are separated by 1$\farcs$23, which has a projected physical separation of 172 AU if at
the nominal distance of Taurus-Auriga ($\sim$140 pc). \label{fig3}}
\end{figure}

\clearpage

\begin{figure}
\epsscale{1.50}
\plotone{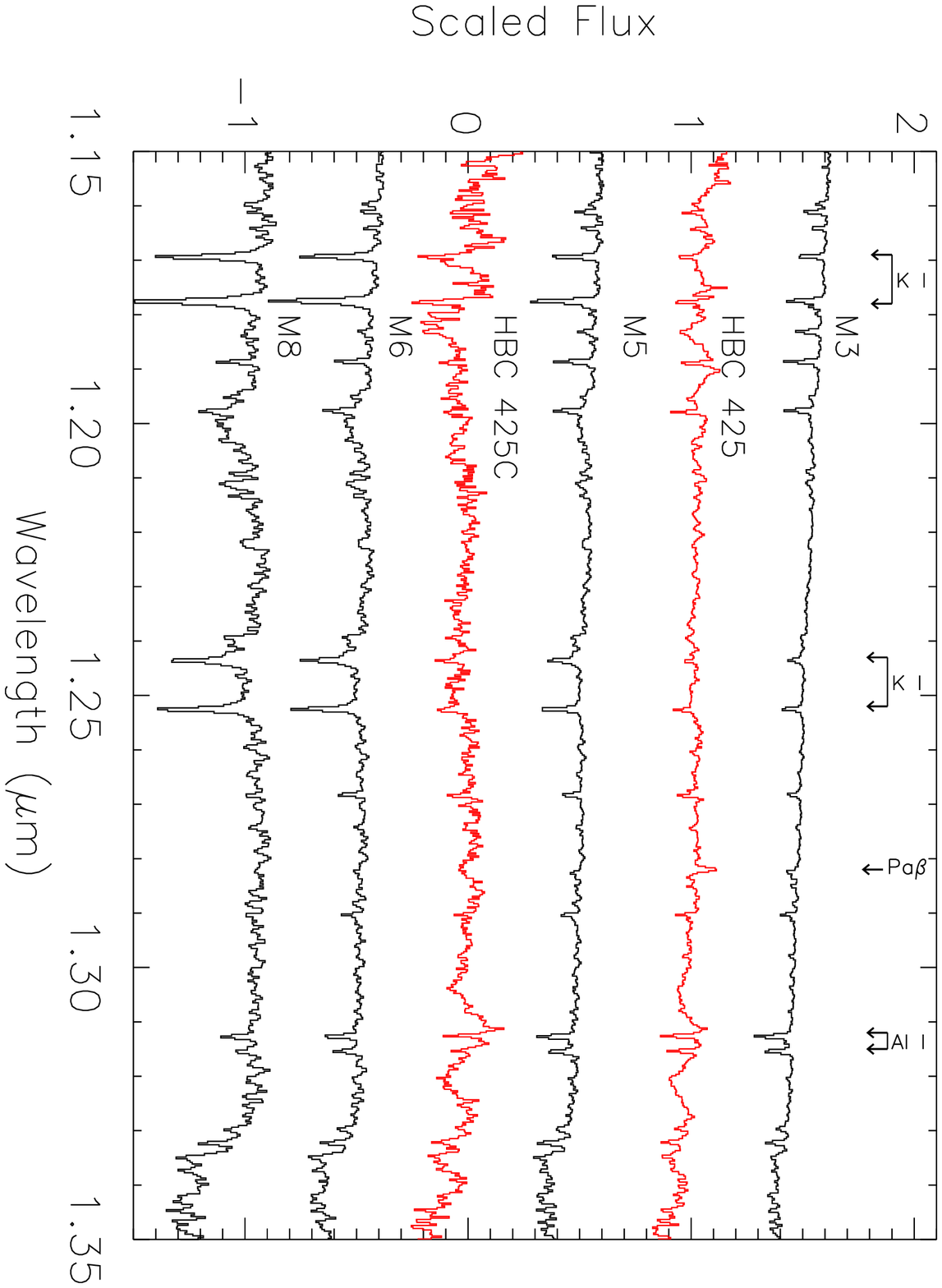}
\caption{NIRSPAO $J-$band spectra of HBC 425 and HBC 425C (in red) plotted with sources spanning 
a range of spectral types obtained from the IRTF spectral library (Rayner et al. 2009). \label{fig4}}
\end{figure}

\clearpage

\begin{figure}
\epsscale{1.50}
\plotone{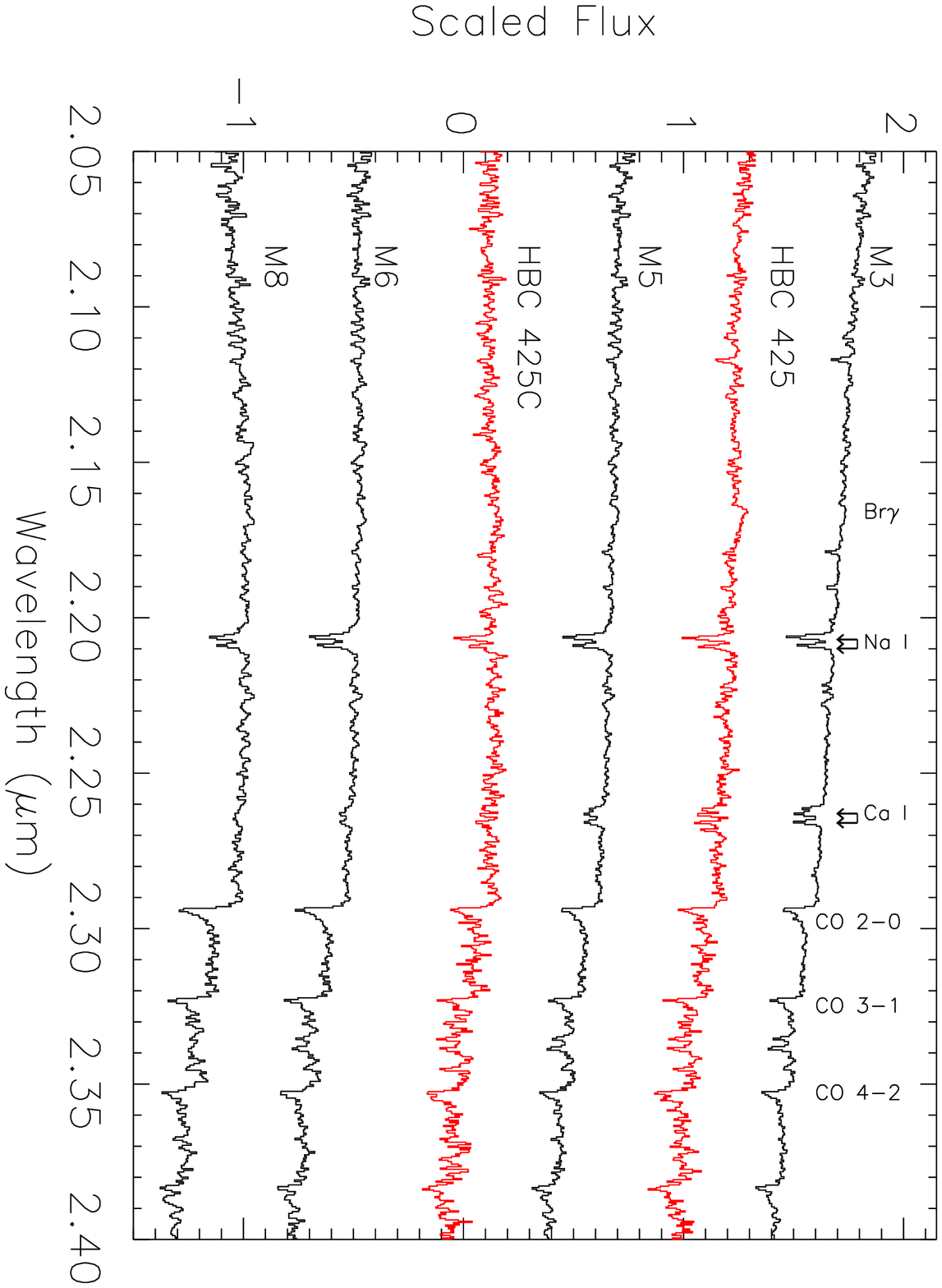}
\caption{NIRSPAO $K-$band spectra of HBC 425 and HBC 425C (in red) plotted with sources spanning
a range of spectral types obtained from the IRTF spectral library (Rayner et al. 2009). \label{fig5}}
\end{figure}

\clearpage

\begin{figure}
\epsscale{1.50}
\plotone{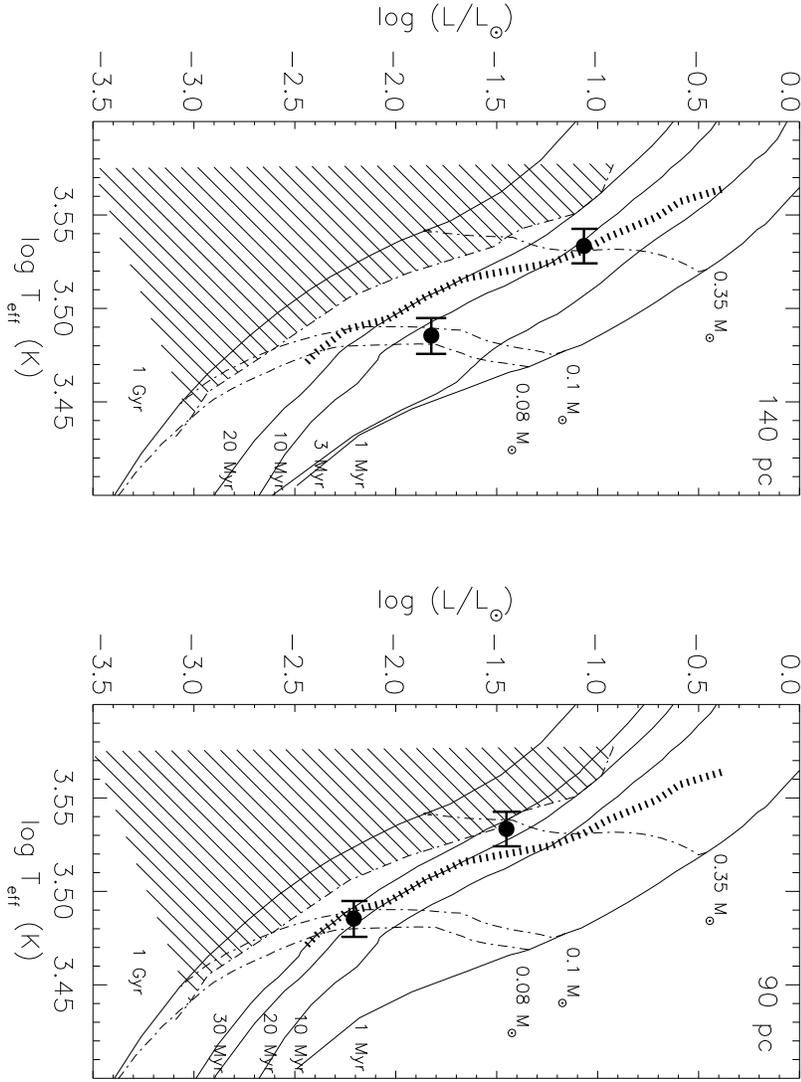}
\caption{(a) The Hertzsprung-Rusell (HR) diagram for HBC 425 and HBC 425C, plotted using the effective temperature
scale of Luhman et al. (2003) and the main sequence bolometric corrections of Kenyon \& Hartmann (1995).
Superimposed are the 1, 3, 10, 20 Myr and 1 Gyr isochrones, and the 0.35, 0.1, and 0.08 M$_{\odot}$ evolutionary tracks of
Baraffe et al. (1998), placed at the nominal distance of Taurus-Auriga $\sim$140 pc. Also shown are 
the theoretical lithium depletion curves of Chabrier \& Baraffe (1997). The onset of depletion is 
represented by the enhanced dashed curve and the cross-hatched region represents significant depletion 
of $\sim$3 orders of magnitude or greater from initial abundance levels. (b) Identical HR diagram
assuming a distance of 90 pc. HBC 425 lies near or within the region of significant lithium depletion
while the onset of depletion should be occurring in HBC 425C.
\label{fig6}}
\end{figure}

\clearpage

\begin{deluxetable}{lccc}
\tabletypesize{\scriptsize}
\rotate
\tablecaption{Near-Infrared Photometry for HBC 425 and HBC 425C \label{tbl-1}}
\tablewidth{0pt}
\tablehead{
\colhead{Source} & \colhead{$J$} & \colhead{$H$} & \colhead{$K_{S}$} 
}
\startdata
HBC 425\tablenotemark{a} &  11.44$\pm$0.02 & 10.83$\pm$0.02 & 10.54$\pm$0.02 \\
HBC 425C\tablenotemark{b} & 13.17$\pm$0.08 & 12.61$\pm$0.06 & 12.23$\pm$0.03 \\
\enddata
\tablenotetext{a}{2MASS photometry for HBC 425 is corrected for the equally luminous spectroscopic binary companion.}
\tablenotetext{b}{Magnitudes for HBC 425C are calibrated using the 2MASS photometry for HBC 425.}
\end{deluxetable}

\clearpage

\begin{deluxetable}{ccccc}
\tabletypesize{\scriptsize}
\rotate
\tablecaption{Observed Properties of HBC 425 and HBC 425C \label{tbl-2}}
\tablewidth{0pt}
\tablehead{
\colhead{Parameter} & \colhead{System}  & \colhead{Primary} & \colhead{Secondary}  & \colhead{HBC 425C}
}
\startdata
Julian Date                           & 2,455,532.925     &                   &              &     \\
W(Ca II K)  (\AA)                     &  ...              &   $-$1.72         &  $-$2.37     &  ...   \\
W(H$\delta$) (\AA)                    &  $-$6.43          &   ...             &   ...        &  ...   \\
W(H$\gamma$) (\AA)                    &  $-$8.96          &   ...             &   ...        &  ...   \\
W(H$\alpha$) (\AA)                    &  $-$27.35         &   ...             &   ...        &  ...   \\
H$\alpha$ 10\% width (km s$^{-1}$)    &  366              &   ..              &   ...        &  ...   \\
W(Li I) (\AA)                         &  ...              &   $\le$0.03       &  $\le$0.03   &  ...   \\
Radial Velocity (km s$^{-1}$)         &  +17.89           &   $-$10.67        &   +46.32     &  ...   \\
\\
\hline
\\
Parameter           & System             & Primary          & Secondary            & HBC 425C \\
\\
\hline
Julian Date                           & 2,455,846.555     &                   &              &             \\
W(H$\gamma$) (\AA)                    &  $-$14.48         &   ...             &   ...        &  ...        \\
W(H$\beta$) (\AA)                     &  $-$14.36         &   ...             &   ...        & $-$5.8      \\
W(H$\alpha$) (\AA)                    &  $-$40.0          &   ...             &   ...        & $-$17.2     \\
H$\alpha$ 10\% width (km s$^{-1}$)    &  462              &   ...             &   ...        & 200:\tablenotemark{a}         \\
W(Li I) (\AA)                         &  ...              &   $\le$0.05       &  $\le$0.05   & 0.32$\pm$0.05\tablenotemark{b}  \\
W(Ca II $\lambda$8542) (\AA)          &  $-$0.58\tablenotemark{c}          &   ...             &   ...        & $-$0.2        \\
Radial Velocity (km s$^{-1}$)         &  ...              &   +16.77          &   +16.77     & +15.9        \\
\\
\hline
\\
Parameter           & System             & Primary          & Secondary            & HBC 425C \\
\\
\hline
Julian Date                           & 2,455,853.454     &                   &              &             \\
W(H$\gamma$) (\AA)                    &  $-$11.52         &   ...             &   ...        &  ...        \\
W(H$\beta$) (\AA)                     &  $-$14.08         &   ...             &   ...        &  ...    \\
W(H$\alpha$) (\AA)                    &  $-$36.83          &   ...             &   ...        &  ...     \\
H$\alpha$ 10\% width (km s$^{-1}$)    &  468              &   ...             &   ...        &  ...         \\
W(Li I) (\AA)                         &  ...              &   $\le$0.05       &  $\le$0.05   &  ...  \\
W(Ca II $\lambda$8542) (\AA)          &  ...          &  $-$0.18\tablenotemark{c}   &   $-$0.32\tablenotemark{c}       & ...        \\
Radial Velocity (km s$^{-1}$)         &  +17.89           &   +41.42          &  $-$5.53     & ...        \\
\enddata
\tablenotetext{a}{Uncertain due to scattered light contamination from HBC 425.}
\tablenotetext{b}{Uncorrected for scattered light contamination. See text for additional details.}
\tablenotetext{c}{Measured from the base of the absorption line profile.}
\end{deluxetable}

\clearpage

\begin{deluxetable}{lcc}
\tabletypesize{\scriptsize}
\rotate
\tablecaption{Inferred Properties for HBC 425 and HBC 425C \label{tbl-3}}
\tablewidth{0pt}
\tablehead{
\colhead{Parameter}   & \colhead{HBC 425 A+B} & \colhead{HBC 425C}
}
\startdata
Mass (M$_{\odot}$)     &     0.35             &    0.09   \\
T$_{eff}$ (K)          &     3415             &    3058   \\
Spectral Type          &     M3               &    M5.5   \\
log L/L$_{\odot}$\tablenotemark{a}       &    $-$1.07           &   $-$1.82 \\
Mass Ratio (q)         &     1.00             &    ...    \\
Semi-major Axis (AU)   &    $<$0.2          &    172\tablenotemark{b}    \\
Period (days)          &    $<$54           &    ...    \\
\enddata
\tablenotetext{a}{Assumes $d$=140 pc, consistent with Taurus-Auriga.}
\tablenotetext{b}{Projected separation assuming $d$=140 pc.}

\end{deluxetable}

\clearpage

\end{document}